\documentclass[12pt]{article}
\usepackage[dvips]{graphicx}
\usepackage{pdproc}

  \makeatletter 
  \def\@cite#1{[#1]} 
  \makeatother    
\begin{document}

\renewcommand{\thefootnote}{\alph{footnote}}

\title{
Equivalence Classes in Gauge Theory
on the Orbifold $S^1/Z_2$
}

\author{Yoshiharu Kawamura}

\address{ 
Department of physics, Shinshu university\\
Matsumoto, 390-8621, Japan
\\{\rm E-mail: haru@azusa.shinshu-u.ac.jp}}

\abstract{
After a brief review of orbifold grand unified theory, 
we discuss two topics related to the choice of boundary conditions
on the orbifold $S^1/Z_2$:
dynamical rearrangement of gauge symmetry 
and equivalence classes of boundary conditions.
}

\normalsize\baselineskip=15pt

\section{Introduction}

The standard model (SM) has been established 
as an effective theory below the weak scale,
but the SM cannot be an ultimate theory of nature 
because of several problems.
Why is the electric charge quantized?
What is the origin of anomaly free sets for matter fields?
It contains many independent parameters.
There is a problem related to the weak Higgs doublet called $\lq$naturalness problem'.
It does not describe gravity.
The situation of the first three problems is improved 
in grand unified theories (GUTs).
The fourth problem is technically solved by the introduction of supersymmetry (SUSY).
The ideas of grand unification and SUSY are so attractive that 
we'd like to go with these ideas.

Grand unification has been paid much attention to as a model beyond the SUSY extension of SM.
Towards the construction of a realistic SUSY GUT, we encounter several problems. 
Why is the proton so stable?
What is the mechanism to break a grand unified symmetry?
There is a problem related to Higgs masses called $\lq$triplet-doublet splitting problem'.
What is the origin of fermion mass hierarchy and mixing?
Many intriguing ideas have been proposed to solve them
by extending the structure of Higgs sector, but the final answer has not been obtained.
Hence we'd like to attack these problems from another angle.

Now it's time to tell you our standpoint and goal.
Our standpoint is that we shall adopt the idea of grand unification and SUSY, and
our goal is to construct a realistic GUT
by introducing extra dimensions.
But we still have a long way to go there, and hence
the goal of my talk is to introduce GUT on the orbifold $S^1/Z_2$ 
and to discuss two interesting topics
related to the choice of boundary conditions (BCs) on the orbifold, 
which will help us in a realistic model building.

\section{Orbifold Grand Unified Theory}

An orbifold is a space obtained by dividing a manifold with
some discrete transformation group, which contains fixed points.\footnote{
Orbifolds were initially utilized for the construction of
4-dimensional heterotic string models.\cite{Orbifold}} 
Fixed points are points that transform into themselves 
under the discrete transformation.
The simple example is $S^1/Z_2$, and 
it is obtained by dividing a circle $S^1$
(whose radius is $R$) with the $Z_2$ transformation, $y \to -y$.
As the point $y$ is identified with the point $-y$,
the $S^1/Z_2$ is regarded as a line segment whose length is $\pi R$.
The both end points $y = 0$ and $\pi R$ are fixed points under the $Z_2$ transformation.

Let us adopt the $\lq$brane world scenario' with the help of orbifold fixed points.
We assume that the space-time is factorized into
a product of 4-dimensional (4D) Minkowski space and the orbifold $S^1/Z_2$.
Their coordinates are denoted by $x$ and $y$, respectively.
Our 4D world is assumed to be sitting on one of the fixed points.
Although the point $y$ is identified with the points
$y + 2 \pi R$ and $-y$, but a value of field does not necessarily take
an identical value at these points.
If we require that the Lagrangian density ${\cal L}$ takes a single-value,
the following BCs are allowed for a field $\Phi(x,y)$,
\begin{eqnarray}
&~& \Phi(x, -y) = T_{\Phi}[P_0] \Phi(x, y) , ~~~
\Phi(x, \pi R -y) = T_{\Phi}[P_1] \Phi(x, \pi R + y) ,
\nonumber \\
&~& \Phi(x, y + 2 \pi R) = T_{\Phi}[U] \Phi(x, y) ,
\label{BC's-Phi}
\end{eqnarray}
where $T_{\Phi}[P_0]$, $T_{\Phi}[P_1]$ and $T_{\Phi}[U]$
represent appropriate representation matrices 
including an arbitrary sign factor.
Those matrices belong to some elements of transformation group of ${\cal L}$ 
and satisfy the relations:
\begin{eqnarray}
T_{\Phi}[P_0]^2 = T_{\Phi}[P_1]^2 = I , ~~
T_{\Phi}[U] = T_{\Phi}[P_0] T_{\Phi}[P_1] ,
\label{Phi-rel}
\end{eqnarray}
where the $I$ stands for a unit matrix.
The eigenvalues of $T_{\Phi}[P_0]$ and $T_{\Phi}[P_1]$
are interpreted as $Z_2$ parities in the fifth direction
for each component of $\Phi$. From the
Fourier-expansion for bulk fields, 
we find that the coefficients are 4D fields and they
acquire quantized masses of $n/R$ $(n = 0, 1, 2, \cdots)$, $n'/R$ and $(2n'-1)/(2R)$ $(n' = 1, 2, \cdots)$
for $Z_2$ parities $(P_0, P_1)= (+1, +1)$, $(-1, -1)$ and $(\pm1, \mp1)$ upon compactification.

Next we discuss a symmetry reduction by BCs on an extra space.
Some symmetries of ${\cal L}$, in general, are broken by imposing different BCs
for each member of a multiplet.\cite{SS}
In our case,
{\it unless all components of $\Phi$ have common $Z_2$ parities, 
a symmetry reduction occurs upon compactification because the zero modes are absent in fields 
with an odd parity.}
This type of symmetry breaking is called the $\lq$orbifold breaking'.

Now we explain the prototype model of orbifold SUSY GUT.\cite{Kawamura}
Our 4D world is supposed to be the 4D hypersurface fixed at $y=0$.
We assume that the 5D bulk fields consist of $SU(5)$ gauge supermultiplets
and two kinds of Higgs hypermultiplets with the fundamental representation.
Three families of matter chiral superfields are located on our 4D brane for simplicity.\footnote{
This assumption is not so strict, but it can be relaxed in the way that some matter superfields
live in the bulk as a member of hypermultiplets.}
With the non-trivial $Z_2$ parity assignment, we obtain the 4D theory
whose zero modes consist of the MSSM gauge supermultiplets 
and two kinds of weak Higgs doublets chiral supermultiplets.
The colored components have an odd parity and no zero mode, 
and hence the triplet-doublet mass splitting is realized elegantly.\footnote{
In 4D heterotic string models,
it has been pointed out that 
extra color triplets are projected out 
by the Wilson line mechanism.\cite{IKNQ}}
In this way, we obtain the just MSSM particles which come from zero modes in the bulk fields
and our 4D brane fields.
The Kaluza-Klein modes don't appear in our low-energy world because they have superheavy masses 
of order $O(1/R)$.
Our simple model has been extended and studied intensively.\cite{Hall&Nomura}

\section{Dynamical Rearrangement of Gauge Symmetry}

The successful realization of mass splitting originates from the non-trivial
assignment of $Z_2$ parities or the choice of twisted BCs.
Here we have an important question:
$\lq$what is the origin of specific parity assignment?'
or $\lq$what is the principle to determine BCs?'
We refer to this problem as the $\lq$arbitrariness problem' and
discuss one aspect of this problem.\cite{HHHK}

There are many kinds of representation matrices that satisfy the relations (\ref{Phi-rel}).
We can find that some of them are gauge equivalent 
because they are related by gauge transformations.
Under the gauge transformation $\Phi(x,y) \to \Phi'(x,y) = T_{\Phi}[\Omega] \Phi(x,y)$,
the BCs can change such that
\begin{eqnarray}
&~& \Phi'(x, -y) = T_{\Phi}[P'_0] \Phi'(x, y) , ~~~
\Phi'(x, \pi R -y) = T_{\Phi}[P'_1] \Phi'(x, \pi R + y) ,
\nonumber \\
&~& \Phi'(x, y + 2 \pi R) = T_{\Phi}[U'] \Phi'(x, y) .
\label{BC's-vphi}
\end{eqnarray}
Here $\Omega$ is a gauge transformation function and operators with a prime are given by,
\begin{eqnarray}
&~& P_0' = \Omega(x,-y) \, P_0 \, \Omega^\dagger (x,y) ~~, ~~
P_1' = \Omega(x,\pi R -y) \, P_1 \, \Omega^\dagger (x,\pi R + y)  ~~, \cr
&~& U' = \Omega(x,y+2\pi R) \,  U \, \Omega^\dagger (x,y) .
\label{BC3}
\end{eqnarray}
Although the symmetry of BCs in one theory differs from that in the other, if two theories
are connected by the BCs-changing gauge transformation,  
they should be equivalent, i.e., $(P_0, P_1, U) \sim (P'_0, P'_1, U')$,
from the gauge principle. 
This equivalence is formally understood as the gauge invariance of the effective potential $V_{\rm eff}$.
Or it is guaranteed by the Hosotani mechanism.\cite{Hosotani} 
We sketch what's going on in the extra dimension.
Let us start with two models which contain
same particle contents but different BCs for them, which are related by some singular gauge transformation.
The physical mass spectrum is determined by both BCs and the vacuum expectation 
value (VEV) of $A_y$.
To obtain the VEV of $A_y$, we need to find the minimum of $V_{\rm eff}$
for each model.
After we find the minimum of $V_{\rm eff}$ and
incorporate the VEV of $A_y$, we arrive at a same mass spectrum for two models.
We refer to this phenomena as the $\lq$dynamical rearrangement of gauge symmetry'.
In this way, theories are classified into equivalence classes of BCs

\section{Equivalence Classes of Boundary Conditions}

In formulating the theory on an orbifold, there are many possibilities for BCs.
We've found that some of them are gauge equivalent and arrive at the concept of
equivalence classes of BCs.
Now the arbitrariness problem is restated as:
$\lq$what is the principle to select a specific equivalence class?'
To obtain a hint as to the problem, we carry out the classification of equivalence
classes and evaluate the vacuum energy density.\cite{HHK}

For the classification of equivalence classes of BCs on $S^1/Z_2$,
we can show that each equivalence class has a diagonal representative for 
$T_{\Phi}[P_0]$ and $T_{\Phi}[P_1]$ 
and the diagonal representatives are specified by three non-negative integers.
{}From this observation, we find that the number of equivalence classes 
is $(N+1)^2$ for $SU(N)$ gauge theories on $S^1/Z_2$.

For the vacuum energy density,
it is given by the value at the minimum of $V_{\rm eff}$.
In our model, the $V_{\rm eff}$ is a function of the background configuration
of gauge field, some numbers which specifies BCs and numbers of species with definite $Z_2$ parities.
As an example, we consider 5D SUSY $SU(N)$ gauge theory 
with the Scherk-Schwarz SUSY breaking mechanism.
We can write down the formula of one-loop effective potential.
(For the formula in the simple case with the vanishing VEV of $A_y$, see Ref.\cite{HHK}.)  
Owing to SUSY, the one-loop effective potential takes a finite value at the minimum 
even after the Scherk-Schwarz SUSY breaking mechanism works.
Hence, if it is allowed,
we can compare with the vacuum energy density among theories that belong to
different equivalence classes.
Many people, however, doubt if the comparison among gauge-inequivalent theories 
is meaningful or not.
We hope that it can make sense in the situation that 
a fundamental theory has a bigger symmetry and BCs are dynamically determined.

\section{Summary}

Orbifold SUSY GUTs possess excellent features and are hopeful.
But we have the problem called the $\lq$arbitrariness problem'.
Using the Hosotani mechanism, we find that theories are classified
into equivalence classes of BCs.
Now the problem is $\lq$what is the principle to select a realistic equivalence class?'
One possibility is a dynamical determination of BCs in a fundamental theory.
Much work would be necessary to solve the problem and to arrive at our goal: the construction of 
a realistic GUT.

\section{Acknowledgements}
This work was supported in part by  Scientific Grants from the Ministry of Education
and Science, Nos. 13135217 and 15340078.

\bibliographystyle{plain}

\end{document}